# Nanoscale characterization of thin immersion silver coatings on copper substrates

Presenting author: Tamás I. Török (*University of Miskolc, Hungary*)
Co-authors: Attila Csik, József Hakl, Kálmán Vad, László Kövér, József Tóth, Sándor Mészáros *(Institute for Nuclear Research Hungarian Academy of Sciences, Debrecen, Hungary)*
Éva Kun, Dániel Sós (*University of Miskolc, Hungary*)

## ABSTRACT

Microelectronic-grade copper foils were immersion silver plated in a home-made non-cyanide alkaline silver nitrate – thiosulfate solution and in two commercially available industrial baths via contact reductive precipitation. The concentration depth profiles of the freshly deposited silver layers were afterwards analyzed at nanoscale resolution by means of Secondary Neutral Mass Spectrometry (SNMS) and Glow Discharge Optical Emission Spectroscopy (GDOES). The thickness of deposited silver layers obtained with 30-60 s immersion time were in the range of 50÷150 nm, depending on the parameters of the immersion procedure. Slight contamination of sulfur from the thiosulfate bath was detectable. Traces of Cr and Na could be observed as well around the interface between the copper substrate and silver deposit. Results also indicate that storage for longer time in air especially at higher than ambient temperatures induces a kind of ageing effect in the deposited layer, changing its composition. In case of samples prepared from home-made solution increasing amounts of copper together with its corrosion products became detectable. The samples were also analyzed by the X-ray photoelectron spectroscopy (XPS) to identify the chemical state of the silver. Nevertheless, working with the two industrial baths we could produce better quality silver layers during the laboratory experiments.

## INTRODUCTION

The quite long history of silver plating both with electrodeposition and via displacement reactions (so-called immersion plating) by using aqueous solutions of some soluble silver compounds started with the cyanides. Though such compounds are rather hazardous, they are still used today for many metal plating tasks even in the microelectronic industries. New trends, however, are aiming at replacing the cyanide anions with other complexing agents (Table 1.) in both electroplating and immersion silver plating. In some cases, the latter technique is getting more favoured as a surface finishing alternative to protect the surface of conducting copper layers from excessive oxidation, for example, before lead-free tin base soldering in many applications utilizing, for instance, printed circuit boards (PCB) makers and users (Table 2.).

**Table 1.** Non-cyanide complexing agents proposed to replace cyanide in silver plating baths [1].

| Name of complexing agent | Year of Citation or the related Patents |
|---|---|
| Thiosulfate | US Patents: 1978, 1994 |
| Hydantion | US Patents: 1997, 2007 |
| Uracil | 2009 |
| Succinimide | US Patents: 1978, 1981 |
| Sulfite | 2010 |
| Ammonia | 1996, 2004 |
| Thiourea | 1998, 1999 |
| HEDTA | 2007 |
| 2-hydroxypyridine | 2009 |
| 5,5-dimethylhydantoin | US Patent: 2012 |
| *Ionic liquids* | 2006, 2007, 2010, 2011 |

**Table 2.** Major surface finishing techniques used in soldering by the microelectronic industry [2].

| Technique | Type of surface treatment | Typical layer thickness |
|---|---|---|
| HASL (Hot Air Surface Leveling) | Hot dip tinning with hot air knife coating thickness control | To cover fully the pads surface for soldering |
| Lead Free HASL | HASL (no lead in the melt) | To cover properly the pads surface for soldering |
| Immersion Tin (Imm Sn) | Chemical replacement reaction during the immersion of copper in the tinning bath | 0.5 … 1.3 μm |
| *Immersion Silver (**ImmAg**)* | *Chemical replacement (contact reduction) of silver while immersion in the bath* | *0.1 ... 0.3 μm* |
| OSP/Entek (Organic Soldering Preservative) | Thin organic protective film (containing antioxidant type inhibitor) | 0.1 … 0.6 μm |
| Immersion Gold (Imm Au) | Chemical replacement reaction in the immersion bath | Au: 0.05 … 0.1 μm<br>Ni: 2.5 … 5 μm<br>(below the gold layer) |
| Hard Gold | E.g. obtained by some Ni addition to the gold electroplating bath | Au: 0.6 … 1 μm<br>Ni: 3.2 … 3.8 μm<br>(below the gold layer) |

For any high quality immersion silver plating procedure the commercially available proprietary ***ImmAg*** baths, in addition to the silver complex, normally might in principle contain several other important additives like buffering agents to control the pH, some corrosion and/or anti tarnishing inhibitors, and also some optional ones, like surface active or wetting agents, grain refiners or brighteners, defoamers, etc. For any successful plating operation there are those technological parameters (temperature, hydrodinamical properties, possible contaminations, etc.) which all should be optimised and controlled meticulously together with the proper and chemically active surface condition of the copper substrate itself.

In view of the above mentioned complexity of the contact reductive precipitation of silver onto copper, in the course of our laboratory experiments, three different immersion plating baths were tested in many laboratory deposition trials. Prepared thin silver deposits were analysed in depth on the nanoscale by means of two powerful surface sputtering elementary analytical techniques, i.e. SNMS (Secondary Neutral Mass Spectrometry) and GDOES (Glow Discharge Optical Emission Spectrometry) in order to reveal the major clues to obtain high quality silver deposits on copper via immersion plating.

## EXPERIMENTAL

### *MATERIALS AND METHODS*

#### *Copper foil*

Thin copper sheets (IMARO/ELTECH) with an average surface roughness of $R_a \approx 0.24$ μm on their shiny side were used after proper cleaning and etching as the copper substrates. The thickness of the copper foils was approximately 90 μm.

#### *Immersion silver plating with the home-made thiosulfate solution (Type A)*

For the preliminary laboratory deposition trials an old and rather short recipe [3] (Type A) was chosen based on which the following experimental procedure was elaborated by us:

1. Surface cleaning
   - degreasing;
   - mechanical scouring to remove surface oxides and any other contaminants possibly attached to the substrate;
   - rinsing with distilled water;
   - sonication (ultrasonic treatment) in acetone for 25 minutes;
   - rinsing with distilled water;
2. Chemical etching
   - etching in 2% $HNO_3$ solution for 3 minutes (max. 7 min);
   - rinsing with distilled water;
   - rapid drying;
3. Silver immersion plating
   - dipping into the immersion silver bath for maximum 1.5 minutes;
4. Washing with water
   - rinsing with distilled water;
   - rapid drying;
   - sonication (ultrasonic treatment) in distilled water for 2 minutes;
   - rapid drying;

For the preparation of the immersion silver bath Type A, 0.75 g silver nitrate ($AgNO_3$) was scaled into a tube, it was filled to about 70 ml with distilled water, stirred permanently by a magnetic stirrer at room temperature. When the silver nitrate dissolved, 10 ml ammonia solution ($NH_3$, 25%) was added. Thereafter, gradually sodium thiosulfate (10.5 g $Na_2S_2O_3$ in total) was mixed into the solution under stirring. Finally, the prepared solution became colorless and transparent.

A complete set of industrial silvering solutions (Type B) was received from a PCB manufacturer and another one was prepared in our laboratory from concentrates purchased from a chemical vendor (Type C). The exact chemical composition of the solution is not known by us.

*Immersion silver plating with industrial solution Type B*

The major laboratory treatment included the following steps:

1. Etching of the copper specimens/substrates in sulfuric acid (~6.5%) containing also hydrogen-peroxide (~5.5%).
2. Pre-dip treatment for surface activation/conditioning (Sterling 2.0 Predip).
3. Immersion plating in a „complexed silver/organic system” (Sterling 2.0 Silver A+B).

Etching time of 30 s at 40°C (104°F; 313.15 K) was found to be optimal with a measured copper surface loss corresponding to about 1.2 μm dissolution. Pre-dipping at ~40°C and silvering at ~50°C lasted somewhat longer (40 s and 75 s, respectively) which parameters were found close to optimal. All the solutions were stirred and we found that rinsing and drying were very important, especially after the silver plating step.

*Immersion silver plating with industrial solution Type C*

All the surface treating solutions (degreaser and oxide remover, surface activator, bright etching, silvering solution, and surface neutralizing and/or antitarnishing solution) were prepared according to the supplier of the chemicals. While working with the Type C solutions mostly the same treatment temperatures and periods were applied during the laboratory plating experiments. Otherwise, the major characteristics of the three immersion plating solutions are summarized in Table 3.

**Table 3.** The three ***ImmAg*** plating solutions used in the laboratory experiments.

| ***ImmAg*** type | Silver (Ag) concentration, g/L | pH | Temp., °C | Complexing agent | Additives |
|---|---|---|---|---|---|
| A | ~ 5 | *Highly alkaline* | *Ambient* | *Thiosulfate & Ammonia* | *Not any* |
| B | *~0.7* | *Acidic: ~ 3.5* | ~50 | *Not disclosed* | *Not known* |
| C | *~1* | *Alkaline: ~9* | *~50* | *Not disclosed* | *Not known* |

*Surface analytical techniques*

By immersion plating one can produce only rather thin (100-300 nm thick) deposits, therefore the elementary concentration depth profiles were measured by GDOES (Glow Discharge Optical Emission Spectrometer, type GD Profiler 2, France) in Miskolc and SNMS (Secondary Neutral Mass Spectrometer, type INA-X from SPECS GmbH, Germany) in Debrecen. In the case of the depth profiles, the horizontal axis represents the sputtering time of argon plasma which is proportional to the crater depth of the specimen. The sputtering time was converted to the depth of the sputtered craters using the profilometers MarSurf M400 MahrSurf SD 26 and Ambios XP-I.

## LABORATORY PLATING EXPERIMENTS AND RESULTS

### *1. Preliminary experiments with thiosulfate bath (Type A)*

During the laboratory experiments many approaches were attempted to develop a sound plating procedure. In our laboratory investigations in the first instance some light yellowish discoloration was often observed on the freshly immersion silver plated specimens (Figure 1) and this discoloration was intensified with increasing time (in a few hours). When the silver plated specimens looked copper-colored, in all the cases, there was some copper present on the surface of the deposited silver detected by the GDOES analysis. Later on the optimization of the process resulted more durable, matt silvery surface. When the cleaning and degreasing of foils were performed by sponge, the silver-treated copper foils became to copper-colored within only after a few hours. Subsequently, we used scouring pad during the cleaning process, whereupon the silver color became durable.

It was also observed, that whenever a properly cleaned copper test foil was dipped into the immersion silver bath very soon (in 1-2 s) a continuous solid silver layer was built up. At the first experiments, the silver layer peeled off because of the long-term dipping in the silver solution. Through repeated experiments the maximum of the peeling free useful dipping time was set finally to 90 seconds.

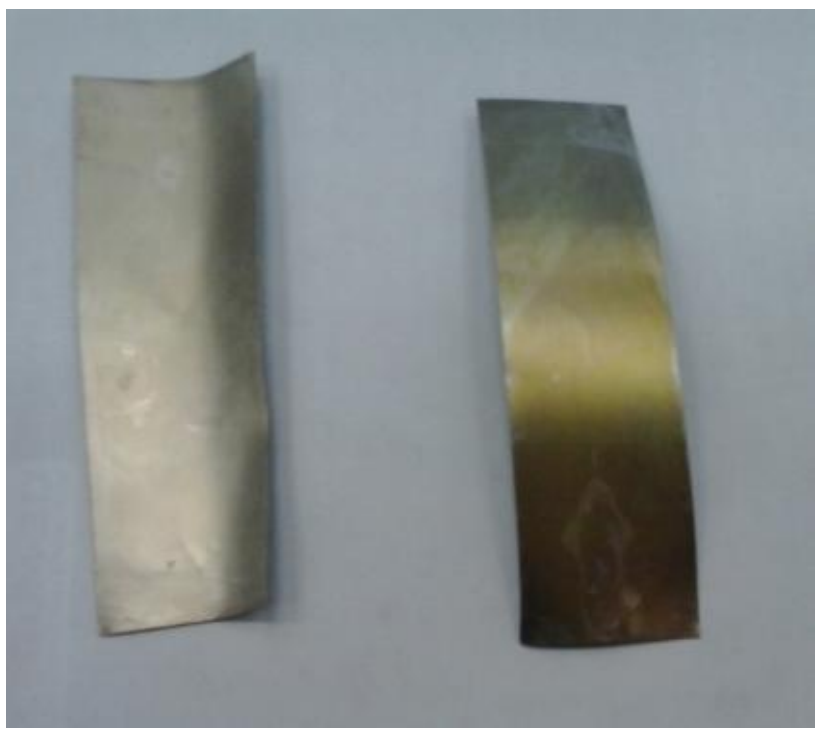

**Figure 1.** Immersion silver plated copper foils prepared without prior scouring.
Left: promptly after coating. Right: one hour later.

To find the reason for the discoloration, we have examined many chemical and/or transport processes that could induce the diffusion of the atoms within layers. We have followed the concentration changes in the silver plated copper specimens during a 3 weeks period (20 min, 1 hour, 2 hours, 4 hours, 1 days, 2 days, 4 days, 1 week, 3 weeks) of ambient annealing. The main curves are shown in Figure 2. It is remarkable, that while the position of copper peaks remained unchanged, the silver layer seems to be diffused into the bulk copper with elapsing by time.

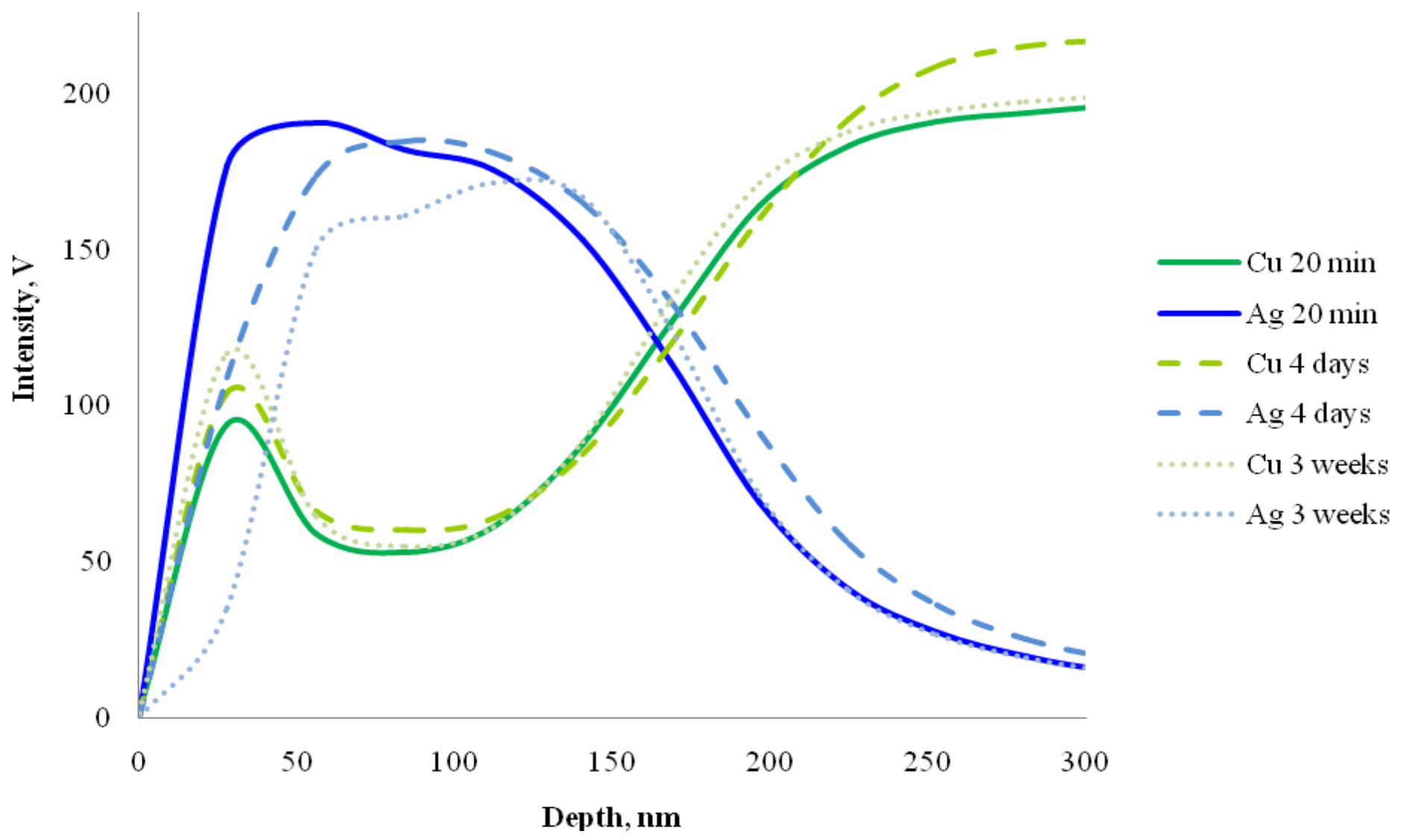

**Figure 2.** GDOES depht profiles of single immersion silver plated copper foil (average of 5 parallel measurements on different samples).

Compared to the GDOS techique the SNMS has lower sputtering rate (~0.1 nm/s) and this provides for us an opportunity for a more detailed analysis of the produced silver layers. The SNMS depth profile of the sample stored in air for 3 weeks was measured and the contaminants were detected in the formed layered structure as well (Figure 3). Sulfur runs together with the Ag, while Na is enriched on the Ag surface and at the Ag/Cu interface. These tendencies are associated with the anionic/cationic nature of the constituents (with respect to Ag) during the immersion plating process. The Ag/Cu surface is partly contamined with oxygen. Oxygen can be found on the Ag surface as well, however its magnitude is partly affected by the instrumental background. The detected Cr originates most probably from the surface treatment of the Cu foils.

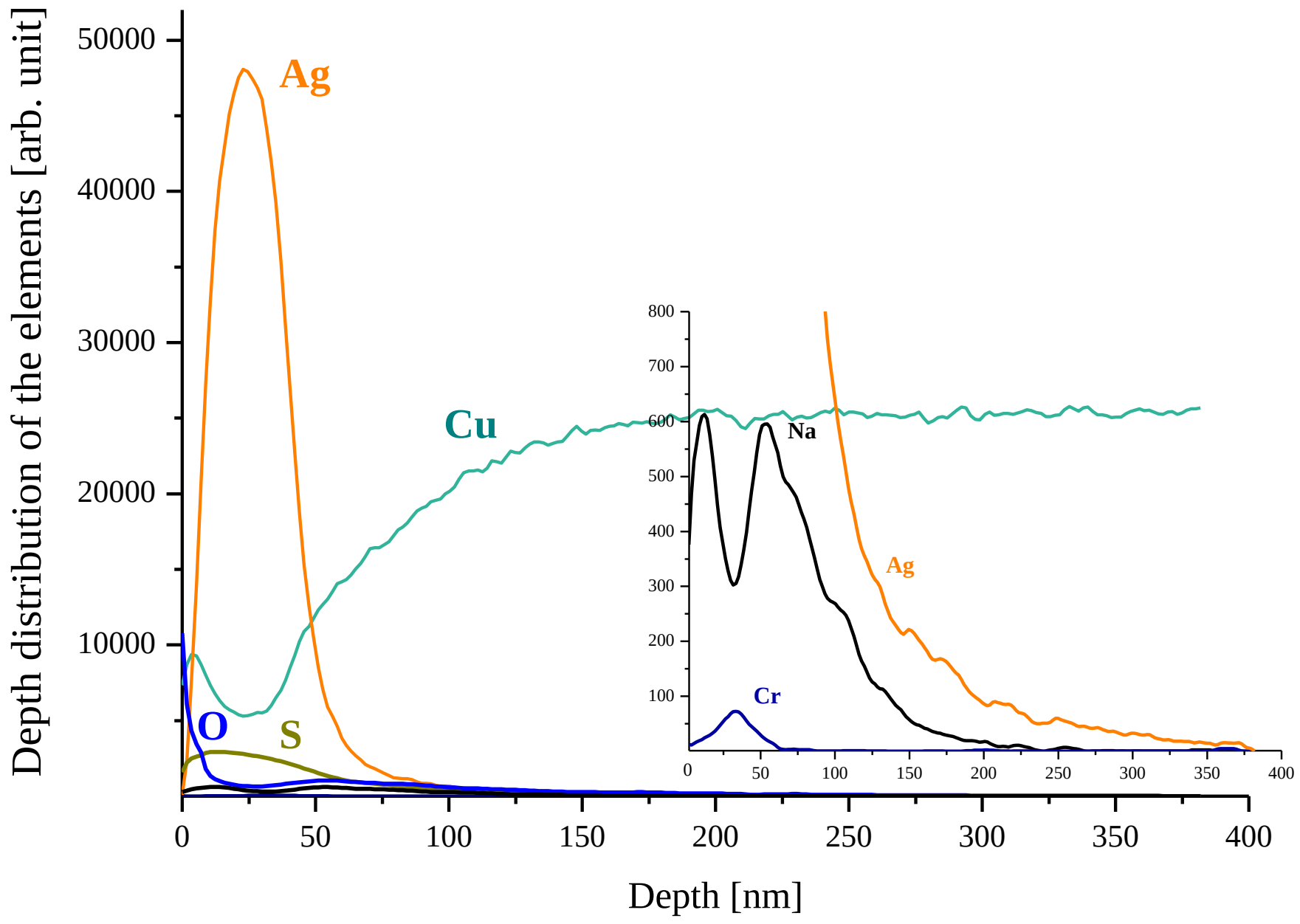

**Figure 3.** SNMS depth profile of a silver plated copper foil stored in air for 3 weeks.

*2. Laboratory plating results with the industrial baths (Type B, C)*

The two industrial immersion plating baths were tested in several sets of laboratory experiments and many of the freshly silver plated copper specimens (foils) were examined via GDOES analysis (Figure 4). Argon sputtering removed the thin silver deposits rather quickly after which the copper substrates were eroded further by the argon plasma until the detected Ag intensities almost diminished and the Cu intensities reached the same plateau. As the substrates were the same material in both case (bath Type B and C) the significantly lower Ag intensities recorded with the Type C specimens indicate structural differences in the silver coatings. This diversity is reflected in the difference of Ag/Cu interface thicknesses, which in turn is connected to the coating surface roughness. As a consequence, type B coating can be regarded as more compact and robust to environmental impacts.

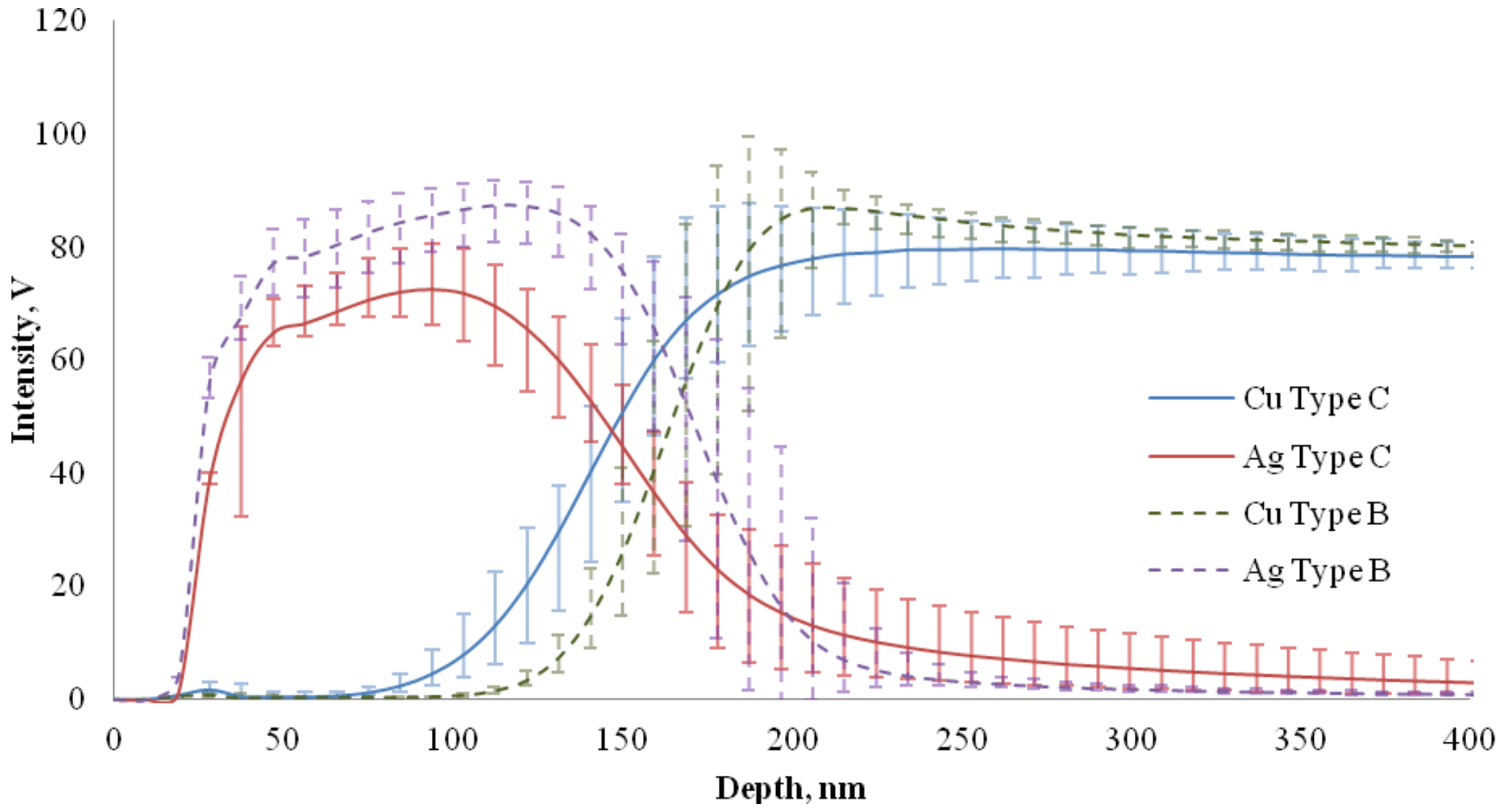

**Figure 4.** GDOES depth profiles of C and B type silver plated specimens (average, minimum and maximum values obtained from ten palallel measurements).

SNMS investigations also were performed for these samples as well. It was found that both silver layers of Types B and C were clean, only a small amount of carbon was be found in the deposited silver layer and accumolation of the Na and S at the Ag/Cu interface (Figure 5). It must be noted, that contrary to the sample prepared from the bath Type A, the sulfure has been found only at the Ag layer/substrate interface and not in the whole Ag layer. No Cr was visible in case of these samples at the Ag/Cu interface.

To confirm these results, X-ray photoelectron spectroscopy (XPS) spectrum was acquired at the depth of 25 nm of the Ag layer (Figure 6). The targeted depth was achieved by sputtering in the SNMS chamber. The sample was subsequently transferred in a common vacuum system into the XPS chamber. Because the SNMS and XPS instruments are connected with each other through the vacuum tube, it was possible to move the sample in a high vacuum preventing of the freshly sputtered surface from contamination. The narrow scan of the $Ag_{3d}$ peak (Figure 7) fits well with the spectra of the reference Ag metal sample. This confirms SNMS results, namely that the silver layer is pure, no contamination exists in the layer (e.g. oxide, sulfate ), the silver is present in metallic state. The binding energies of the silver 3d lines agree in ±0.1 eV with the NPL (UK) and NIST (USA) standard values of 368.25 eV. The energy difference between the Ag 3d and the Ag MNN Auger-lines agree in ±0.1 eV as well with the mentioned standard values. Both of the measurements (sample and standard) of the Ag 3d lines were made in the same instrumental set up of the XPS machine. During the XPS measurements the vacuum level was in the $10^{-10}$ mbar range.

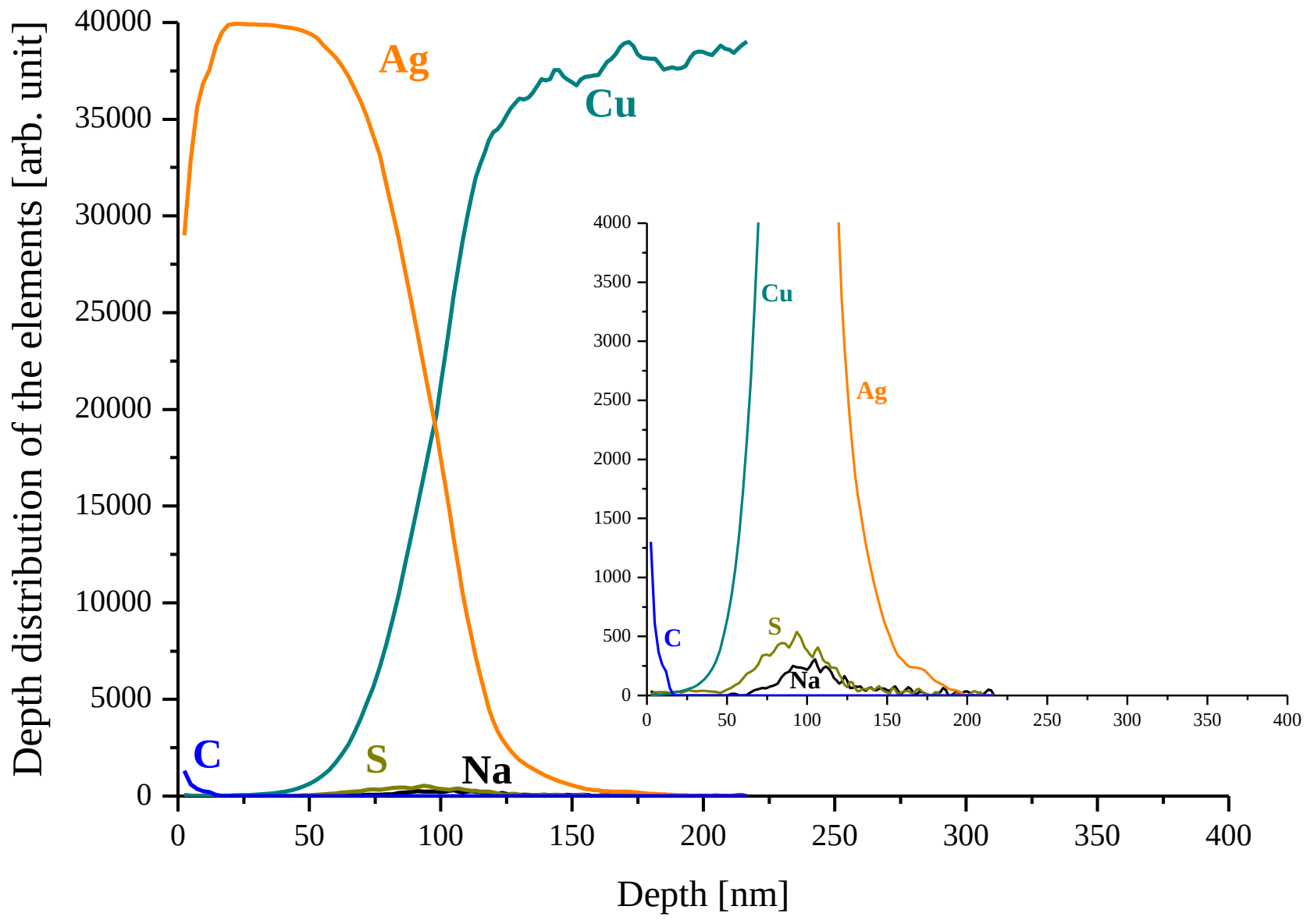

**Figure 5.** SNMS depth profile of one of the specimens coated using the Type C silver plating method.

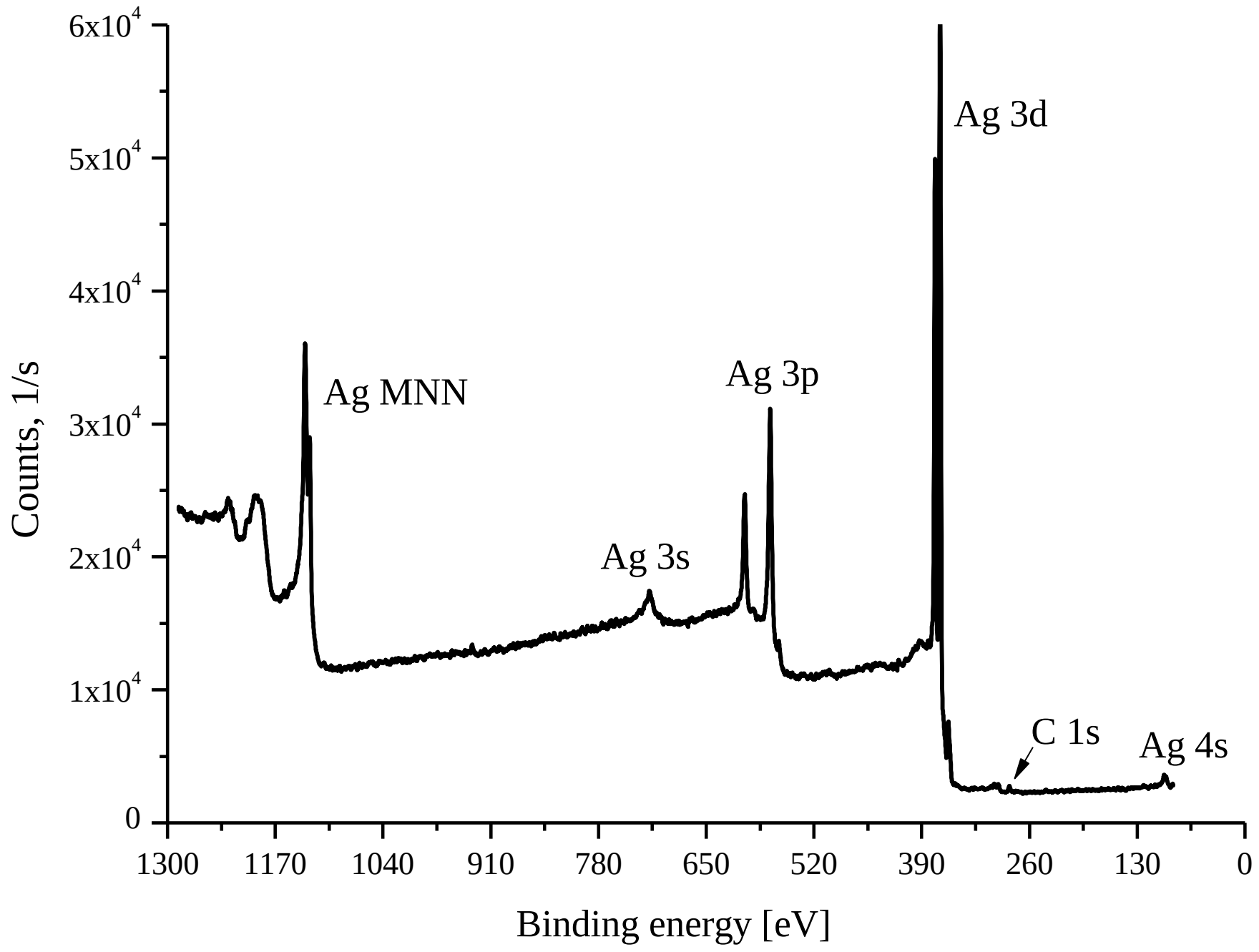

**Figure 6.** XPS survey scan recorded at 25 nm depth in the silver coating of the same Type C specimen as in Figure 5.

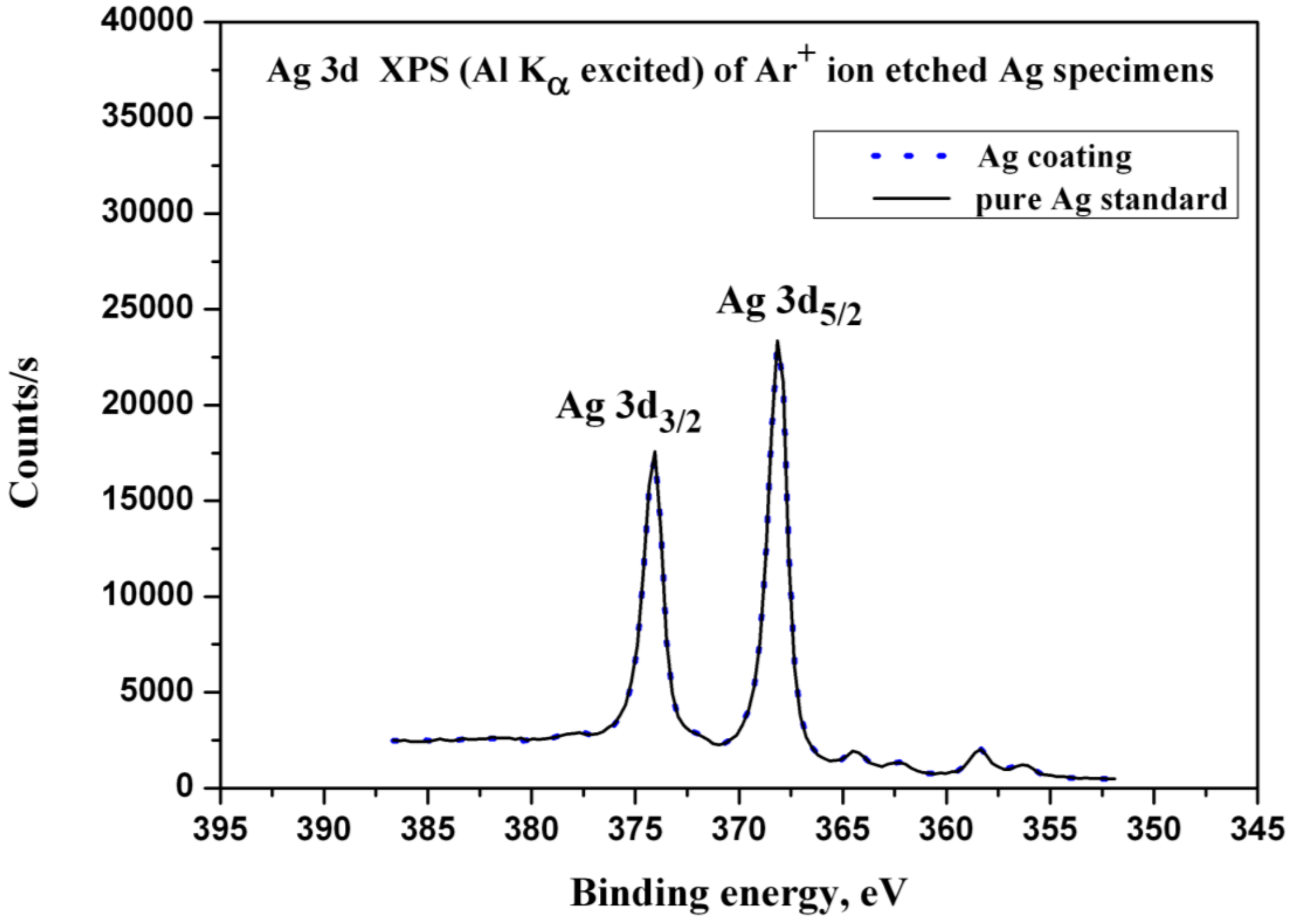

**Figure 7.** Comparison of the XPS narrow scans of the Ag 3d peaks measured in the case of the silver coating and the reference pure silver samples.

## DISCUSSION

At first glance, immersion silver plating might be thought of as a quite simple process to cover the surface of pure copper substrates (e.g. pads of PCBs). In principle, it works via a simple chemical displacement reaction between the solid Cu and $Ag^+$ cations of the plating solution. Such solutions should at least contain a soluble salt of silver and some necessary additives to control, first of all, the kinetics of the contact reductive precipitation of silver onto the copper substrate which is concurrently dissolving into the plating bath. Therefore, in reality, there are many parameters which have to be controlled during the plating process in order to produce an even and sound (nonporous) layer with good adherence. All these facts explain why even the basic chemical constituents of the immersion silver plating solutions are still under permanent development [4].

For our laboratory immersion silver plating experiments, first we have chosen an alkaline non-cyanide silver plating ‘recipe’ with very limited accessible information [3], based on which almost the whole immersion plating methodology had to be developed during our laboratory trials. But, eventually several sets of thin silver plated specimens were successfully prepared, although many of the silver coated specimens showed some kind of discoloration. It is well known that the number of causes of colour of objects as perceived by the human eyes can be rather high, i.e. around fifteen [5]. In the case of pure silver metal, its light absorption peak lies in the ultraviolet region, and as a result, silver maintains high reflectivity evenly across the visible spectrum and we see it as a pure white [6]. However, tiny globular silver grains, produced for instance in transparent polyacrilamide gels, which contained different proteins as well, appeared blue (with grain diameters of 40-100 nm) then yellow (21-39 nm) or brown (17-35 nm), which peculiar phenomena were described in detail already in 1988 [7]. Since then, the colouring effects of metal nanoparticles, for example the plasmon resonance mechanism, has generated much attention [8].

During our laboratory experiments, when we first noticed the discoloration of our immersion silver coated copper specimens stored in air, the above mentioned aspects of the formation of surface colors were all quite helpful in trying to explain the situation in our case. The surface corrosion (often called tarnishing) of silver, especially in contaminated air, is also well known among metal processors, silversmiths and art collectors [9], and the thermodynamic stabilities of the most common corrosion products of silver and copper, in principle, confirm those observations (Figure 8). Conserning the sulfur-copper reaction and taking into account the results of SNMS depth profiling (Figure 3.) we can conclude that the this cause colour changes of the sample prepared from the bath Type A.

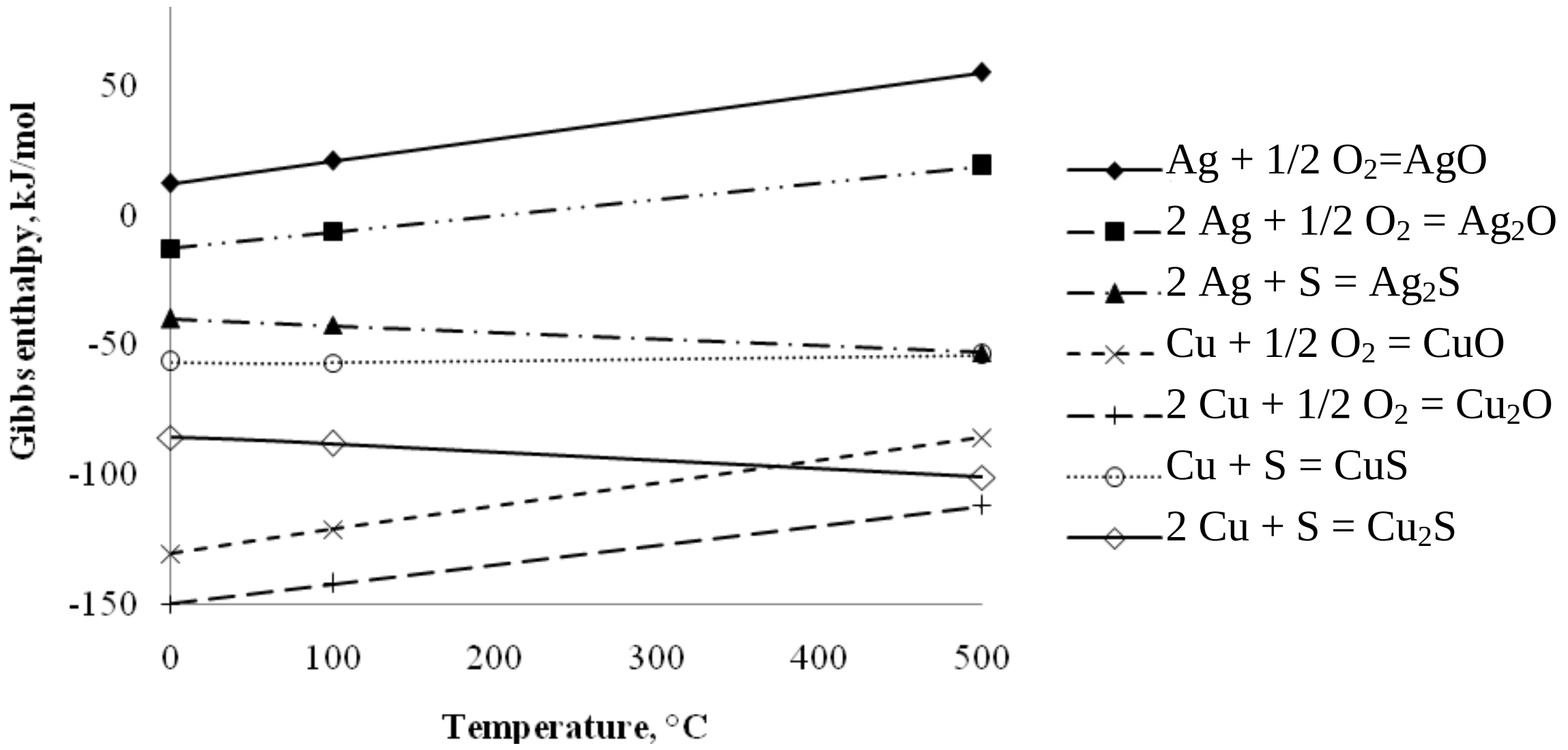

**Figure 8.** Standard Gibbs enthalpies of formation for the oxides and sulfides of silver and copper, respectively, as a function of the temperature.

It is seen from the data (calculated by the HSC Chemistry 5.0 Thermodynamic Database and Software) given in Figure 7, that both oxides of copper are quite stable, while silver should form sulfides more readily than oxides. And, when the more noble silver metal is in contact with copper, like in our case, there is also an additional electrochemical driving force towards the anodic oxidation of copper. Moreover, if the circumstances are favorable (e.g. in wet conditions) the galvanic corrosion of copper should occur first. However, modern silver plating baths and procedures are so well developed - as we could see it while testing two industrial baths in our laboratory experiments - that users should not worry much about it provided all the pre-treatment and other procedures and technological parameters are controlled well.

## CONCLUSIONS

Non-optimized simple immersion silver plating bath like our home-made non-cyanide and highly alkaline one was found not suitable for high-tech industrial applications, although we could gain quite useful information about the weak points of immersion silver plating on copper during our preliminary experiments. First, the rather thin (around 110÷150 nm) and porous silver deposits did not last long without surface discoloration, partially due to the incorporation of sulfur from the thiosulfate bath itself and migration of the constituting Ag and Cu atoms. Second, proper surface cleaning and bright etching were found also crucial steps prior to develop a sound and smooth silver deposit. Using new complexing agents and better additives in the newer industrial immersion silver baths, there should be much less concern that ***ImmAg*** bath were inappropriate or less advantageous than gold for example in high-tech applications even in the microelectronic industries.

## ACKNOWLEDGEMENT

The research work presented in this paper based on the results achieved within the TÁMOP-4.2.1.B-10/2/KONV-2010-0001 project and carried out as part of the TÁMOP-4.2.2.A-11/1/KONV-2012-0019 project in the framework of the New Széchenyi Plan. The realization of this project is supported by the European Union, and co-financed by the European Social Fund.